\documentclass[runningheads]{llncs}

\usepackage[T1]{fontenc}
\usepackage{graphicx}
 \usepackage[misc]{ifsym}
 
\usepackage{mathtools}
\usepackage{oz}
\usepackage{enumerate}
\usepackage{array}
\usepackage{xcolor}

\newcommand{\df}[1]{#1_{def}}

\newcommand{\wps}{wp_s}
\newcommand{\wpstl}{wp_{STL}}
\newcommand{\wppsf}{wp_{PSF}}

\renewcommand{\rename}[2]{[#1\backslash #2]}

\newcommand{\calL}{{\cal L}}

\newcommand{\sequ}{\,;\hspace*{-0.3ex}}
\newcommand{\skipc}{\text{\small\sf skip}}
\newcommand{\fence}{\text{\small\sf fence}}
\newcommand{\leak}[1]{\text{\small\sf leak }#1}

\newcommand{\ifc}{\text{\small\sf if\,}}
\newcommand{\elsec}{\text{\small\sf \,else\,}}
\newcommand{\whilec}{\text{\small\sf while\,}}

\newcommand{\base}{{\flat}}

\definecolor{mygray}{gray}{0.95}
\newcommand{\codebox}[1]{\colorbox{mygray}{#1}}

 \usepackage{url}

  \makeatletter
  \AtBeginDocument{%
  \def\doi#1{\url{https://doi.org/#1}}}
  \makeatother

\begin{document}

\title{Detecting speculative data flow vulnerabilities using weakest precondition reasoning} 
\titlerunning{Detecting speculative data flow vulnerabilities}
\author{\vspace{1cm}}
\institute{}
\author{Graeme Smith
\orcidID{0000-0003-1019-4761}}
\institute{Defence Science and Technology Group, Australia\\
School of Electrical Engineering and Computer Science,\\ The University of Queensland, Australia\\
\email{g.smith1@uq.edu.au}}

\maketitle


\begin{abstract}
Speculative execution is a hardware optimisation technique where a processor, while waiting on the completion of a computation required for an instruction, continues to execute later instructions based on a predicted value of the pending computation. It came to the forefront of security research in 2018 with the disclosure of two related attacks, Spectre and Meltdown. Since then many similar attacks have been identified. While there has been much research on using formal methods to detect speculative execution vulnerabilities based on predicted control flow, there has been significantly less on vulnerabilities based on predicted data flow. In this paper, we introduce an approach for detecting the data flow vulnerabilities, Spectre-STL and Spectre-PSF, using weakest precondition reasoning. We validate our approach on a suite of litmus tests used to validate related approaches in the literature. 
\end{abstract}

\section{Introduction}
\label{sec:intro}

Modern processors liberally employ speculative execution of instructions to optimise performance. Instructions can be executed before earlier instructions in a program based on predictions of the outcomes of the earlier instructions. The intention is to use latent processing cycles rather than waiting for the completion of computations required for the earlier instructions. When a prediction is found to be correct, the speculatively executed instructions are committed to memory. When a prediction is found to be incorrect, the speculatively executed instructions are rolled back, and execution restarted according to the actual outcome. 

While the rollback of incorrect speculation maintains a program's functionality, traces of speculative execution are left in the processor's micro-architecture and can be exploited by an attacker to gain access to otherwise inaccessible (and hence potentially sensitive) data. The best known such attack, Spectre variant 1 (also known as Spectre-PHT) \cite{koc19}, takes advantage of the pattern history table (PHT), a micro-architectural component used to predict the outcome of a branch instruction. After finding a suitable {\em gadget\/} (i.e., code pattern) in a victim program, an attacker can train the PHT to expect a particular outcome and then use this to exploit the gadget. For example, an attacker could train the PHT to expect the following gadget to execute the body of the if statement. Then, by providing a value of $x$ greater than $array1\_size$, a value beyond the end of $array1$ is accessed in the if statement's body.  This value is subsequently used to access a particular index of $array2$, reading the value at that index into the cache. After rollback, this index can be deduced by a timing attack on the cache  \cite{liu15}. Note that 512 corresponds to the cache line size in bits allowing the attacker to determine, from the affected cache line, the value $array1[x]$.

\[r0 := x;\\
r1 := array1\_size;\\
\ifc (r0 < r1) \{\\
\t1 r1 := array1[r0];\\
\t1 r2 := array2[r1*512];\\
\}
\]

Since the disclosure of such attacks in 2018, a number of formal methods-based approaches for detecting vulnerable gadgets have been developed \cite{cau22}. These have mostly focused on attacks exploiting speculation of control flow in a program, such as Spectre-PHT.
Significantly less work exists on attacks exploiting speculation of data flow such as Spectre variant 4 (Spectre-STL)
\cite{can19}\footnote{Originally described by Jann Horn at \url{https://project-zero.issues.chromium.org/issues/42450580}.}. This attack exploits the incorrect prediction that a load is not dependent on an earlier store and hence can be executed first, missing the Store-To-Load (STL) dependency. 
A similar attack, Spectre-PSF
\cite{cau20,pon22}, relies on the processor incorrectly predicting that a load {\em will\/} depend on an earlier store and speculatively executing the load using the store's value, referred to as Predictive Store Forwarding (PSF). 

In this paper, we provide a weakest precondition-based approach to detecting the data flow Spectre variants, Spectre-STL and Spectre-PSF, building on a recent approach for Spectre-PHT \cite{cou24}. The existing approach is detailed in Section~\ref{sec:wp}. In Section~\ref{sec:data}, we describe the data flow variants of Spectre with simple examples before presenting our formal approach to their detection in Sections~\ref{sec:stl} and~\ref{sec:psf}. In Section~\ref{sec:related} we discuss related formal approaches before concluding in Section~\ref{sec:con}.

\section{Background}
\label{sec:wp}

Winter et al.\ \cite{win21} present an information flow logic based on the weakest precondition (wp) reasoning of Dijkstra \cite{dij76,dij90}. The logic introduces additional proof obligations to standard wp rules to ensure a form of {\em non-interference\/} \cite{gog82}: the proof obligations fail when sensitive information can be leaked to publicly accessible variables or through observation of control flow. 

This logic forms the basis of the approach to detecting Spectre-PHT vulnerable code by Coughlin et al.\ \cite{cou24}. Following \cite{col19}, that approach employs the notion of a speculative context to track the effects of speculative execution. This is incorporated in a weakest precondition transformer $\wps$ which operates over \textit{pairs} of predicates $\langle Q_s, Q\rangle$. The predicate $Q_s$ represents the weakest precondition at that point in the program, assuming the processor is speculating, and $Q$ the weakest precondition when it is not speculating. The security of a program ultimately depends on the non-speculative predicate $Q$ holding in the program's initial state: proof obligations from the speculative state $Q_s$ are taken into account by being transferred to the non-speculative state at points in the program where speculation can begin. 

The rules of $\wps$ are defined over a high-level programming language representing assembly programs. The syntax of an instruction, $\alpha$, and a program, $p$, is defined as follows.

\[\alpha \ddef \skipc | r := e | r := x | x := e | \fence | \leak{e} \\
p \ddef \alpha | p \sequ p | \ifc (b) \{ p \} \elsec \{p\} | \whilec (b) \{p\} \]

\noindent where $r$ is a register, $x$ is a local or global variable (i.e., a memory location which in this paper can be an array access of the form $a[e]$), $b$ a Boolean condition and $e$ an expression. Both $b$ and $e$ are in terms of registers and literals only, as in
assembly code. The language includes a $\fence$ instruction which prevents reordering of instructions (in the context of a processor's memory model) and also
terminates current speculative execution. A special \emph{ghost} instruction\footnote{A ghost instruction is not part of
  the actual code and is used for analysis purposes only.} $\leak{e}$ is inserted into a program to indicate that the
following instruction(s) are part of a gadget that leaks the value $e$ through a micro-architectural side channel when executed
(speculatively, or otherwise).

Before analysing a program with the logic, $\leak$instructions are inserted for each gadget of interest during a pre-pass over the code. Since typical gadgets can be detected syntactically, this is a straightforward task to mechanise. The expression $e$ of the inserted $\leak$instruction is based on what information leaks when the gadget is used in an attack. For the example of Section~\ref{sec:intro}, $\leak r1$ would be inserted immediately above the access to $array2$. After this pre-pass, the code is analysed using the logic to determine whether the information leaked is possibly sensitive and hence the gadget causes a security vulnerability. Since the pre-pass can be customised for different gadgets, the overall approach can be adapted to a variety of attacks, including new attacks as they are discovered.

\subsection{Rules of $\wps$}

{\bf Skip} A $\skipc$ instruction does not change the $\lseq Q_s, Q\rseq$ tuple.

\[\wps(\skipc, \lseq Q_s, Q\rseq) = \lseq Q_s, Q\rseq\]

\noindent{\bf Register update} For each register or variable $v$ in a program, the logic includes an expression $\Gamma_v$ which evaluates to the {\em security level\/} of the information held by the variable. The possible values of security levels form a lattice $(L, \sqsubseteq)$ where each pair of elements $a,b \in L$ has a \emph{join}, i.e., least upper bound, denoted by $a \sqcup b$, and a \emph{meet}, i.e., greatest lower bound, denoted by $a \sqcap b$. The rule for updating a register $r$ to the value of an expression $e$ updates both $r$ and $\Gamma_r$ as follows, where $\Gamma_E(e) = \zBig\sqcup_{r \mem regs(e)}\, \Gamma_r$ is the join of the security levels of the registers, $regs(e)$, to which $e$ refers.

\[\wps(r := e, \lseq Q_s, Q\rseq) = \lseq Q_s\rename{r, \Gamma_r}{e, \Gamma_E(e)}, Q\rename{r, \Gamma_r}{e, \Gamma_E(e)}\rseq\]
where $Q\rename{x_1,...,x_n}{y_1,...,y_n}$ replaces each free occurrence of $x_i$ (for $1 \leq i \leq n$) in $Q$ with $y_i$.\medskip

\noindent{\bf Load} Each variable $x$ has a programmer-defined {\em security policy\/} $\calL(x)$ denoting the highest security level that $x$ may hold. This level may vary as the program executes \cite{mur16,mur18} and hence $\calL(x)$ is an expression in terms of other variables. For example, $\calL(x) = ({\sf if\,} y=0 {\sf \,then\,} secret {\sf \,else\,} public)$, where $secret, public \mem L$, captures that variable $x$ may hold $secret$ information when $y=0$ and $public$ information otherwise. 

 When loading the value of a variable $x$ into a register $r$, it is possible that the security level of that value is undefined, e.g., when it has been set to an input value. Hence, $\Gamma_r$ in the non-speculative state $Q$ is updated to the meet of $\Gamma_x$ and its maximum possible value, $\calL(x)$.  

In the speculative state $Q_s$, $r$ and $\Gamma_r$ are updated with values from memory (referred to as the {\em base state\/} and denoted with a $\base$ superscript) when $x$ is not defined in the speculative context, i.e., an earlier store to $x$ has not occurred, and $x$ and $\Gamma_x$ otherwise. 
This is required to support concurrency since, during speculative execution, another thread may change a value in memory (the base state) but cannot change the corresponding value in the speculative state. Hence, values in the base state and speculative state can differ. The superscripts avoid the base state variables being affected by speculatively executed assignments. 

Whether or not a variable $x$ has been defined in the speculative context is captured by a Boolean ghost variable $\df{x}$. 

\[\wps(r := x, \lseq Q_s, Q\rseq) =\lseq \M (\df{x} \implies Q_s\rename{r, \Gamma_r}{x, \Gamma_x}) \land\\
(\lnot\! \df{x}\implies Q_s\rename{r, \Gamma_r}{x^\base, \Gamma_{x^\base} \sqcap \calL(x)\rename{var}{var^\base}}),\\
Q\rename{r, \Gamma_r}{x, \Gamma_x \sqcap \calL(x)}\rseq\O\]
 where $var$ is the list of program variables, and $var^\base$ the same list with each element decorated with a $\base$ superscript. Note that when $\df{x}$ holds, we can use $\Gamma_x$ directly, rather than the meet with $\calL(x)$. \medskip
  
\noindent{\bf Store} A store to a variable, $x := e$, sets $\df{x}$ to true and replaces each occurrence of variable $x$ and $\Gamma_x$ with expression $e$ and security level $\Gamma_E(e)$, respectively, in both $Q_s$ and $Q$. Additionally, in the non-speculative case non-interference is ensured by checking that

\begin{enumerate}[(i)]
\item the security level of $e$ is not higher than the security classification of $x$, and
\item since $x$'s value may affect the security classification of other variables, for each such variable $y$, 
$y$'s current security level $\Gamma_y \sqcap \calL(y)$  does not exceed its updated security classification when $x$ is set to $e$.
\end{enumerate}
Such checks are not required in the speculative case since, while speculating, values are not written to shared memory.

\[\wps(x := e, \lseq Q_s, Q\rseq) = \lseq \M Q_s\rename{x, \Gamma_x, \df{x}}{e, \Gamma_E(e), true},\\
Q\rename{x, \Gamma_x}{e, \Gamma_E(e)} \land
\Gamma_E(e) \sqsubseteq \calL(x) \land\\
(\all y \cdot \Gamma_y \sqcap \calL(y) \sqsubseteq \calL(y)\rename{x}{e})\rseq\O\]

\noindent{\bf Fence} The $\fence$ instruction terminates any current speculative execution. Hence, any proof obligations in the speculative state
beyond the fence do not need to be considered at the point in the program where a fence occurs. $Q_s$ is therefore replaced by $true$ and $Q$ is unchanged.

\[\wps(\fence, \lseq Q_s, Q\rseq) = \lseq true, Q\rseq\]

\noindent{\bf Leak} The instruction $\leak{e}$ leaks the value of expression $e$ via a micro-architectural side channel, introducing a proof obligation into
both $Q_s$ and $Q$.
\[\wps(\leak{e}, \lseq Q_s, Q\rseq) = \big\lseq Q_s \land \Gamma_E(e) = \bot, \,Q \land \Gamma_E(e) = \bot\big\rseq\]
where $\bot$ denotes the lowest value of the security lattice. Requiring that the leaked information is at this level
ensures that the attacker cannot deduce anything new from the information, regardless of the level of information they
can observe.\medskip

\noindent{\bf Sequential composition} As in standard wp reasoning, sequentially composed instructions transform the tuple one at a time.

\[\wps(p_1 \sequ p_2,  \lseq Q_s, Q\rseq) = \wps(p_1, \wps(p_2, \lseq Q_s, Q\rseq))\]

\noindent{\bf If statement} In the case of Spectre-PHT, speculation can begin at an if statement. Hence, it is at this point in the reasoning that the speculative proof obligation manifests itself as a proof obligation in the non-speculative state. For ease of presentation, we assume that the guard $b$ does not change during speculation, hence the speculative proof obligation can be evaluated in the context of the guard.\footnote{An alternative rule that does not require this assumption is provided in \cite{cou24}.} The speculative proof obligation is from the opposite branch to the one that should be executed, with each variable $\df{x}$ set to false (leaving just the predicates in terms of base variables) and all $\base$ subscripts removed (to identify these base variables with variables in the non-speculative state). 

There is an additional proof obligation $\Gamma_E(b) = \bot$ on the non-speculative state since, in concurrent programs, the value of $b$ can readily be deduced using timing attacks (even when the statement's branches do not change publicly accessible variables) \cite{mur18,smi19}. An if statement might occur within a speculative context (when nested in or following an earlier if statement, for example). The branch that is followed speculatively is, in general, independent of that actually executed later. Hence, the speculative proof obligations from both branches are conjoined to form the speculative precondition.

Given $\lseq Q_{s1}, Q_1\rseq = \wps(p_1, \lseq Q_s, Q\rseq)$ and $\lseq Q_{s2}, Q_2\rseq =  \wps(p_2, \lseq Q_s, Q\rseq)$, we have

\[\wps(\ifc(b)\{p_1\}\elsec\{p_2\}, \lseq Q_s, Q\rseq) =\\
\t1 \lseq\, \M Q_{s1} \land Q_{s2},\M
 (b\implies Q_1 \land Q_{s2}\rename{var^\base, d_1, ..., d_n}{var, false, ..., false}) \land\\
  (\lnot b\implies Q_2 \land Q_{s1}\rename{var^\base, d_1, ..., d_n}{var, false, ..., false}) \land\\
  \Gamma_E(b) =\bot\,\rseq\, .\O\O\]
where $d_1,...,d_n$ is the list of ghost variables of the form $\df{x}$.\medskip

\noindent{\bf While loop} Speculation can also begin at each iteration of a while loop. Similarly to standard wp reasoning, we can soundly approximate the weakest precondition of a loop by finding invariants which imply our speculative and non-speculative postconditions. As with the if rule, a proof obligation
$\Gamma_E(b) = \bot$ must hold in the non-speculative case.

\[\wps(\whilec(b)\{p\}, \lseq Q_s, Q\rseq) = \lseq Inv_s,\, Inv \rseq\]
where $Inv_s \implies Q_s$, $Inv \implies \Gamma_E(b)=\bot \land Inv_s\rename{var^\base,d_1,...,d_n}{var,false,...,false}$ and $Inv \land \lnot b\implies Q$, and given $\wps(p, \lseq Inv_s, Inv\rseq) =\lseq P_s, P\rseq$, then $Inv_s \implies P_s$ and $Inv \land b\implies P$. Like the if rule, the while rule copies the proof obligations in the speculative precondition to the non-speculative precondition, and maintains those in the speculative precondition in case the loop is reached within an existing speculative context.

\subsection{Using $\wps$}
\label{sec:wpsrg}

The property that $\wps$ verifies, when the calculated weakest precondition of a program holds, is {\em value-dependent non-interference\/} based on the definition in \cite{mur16}. This property states that, given two initial states $s_1$ and $s_2$ which agree on the values of variables which are non-sensitive, after executing a prefix of instructions $t$ of the program on each state, the resulting states will continue to agree on the values of variables which are non-sensitive. In other words, the values of variables which are sensitive have no effect on those that are non-sensitive (and hence the sensitive values cannot be deduced from observations of the non-sensitive values). Formally, given a program $c$ with precondition $P$ and postcondition $Q$\footnote{$\Rrightarrow$ denotes logical entailment and binds less tightly then implication ($\implies$).}

\[P \implies \wps(c,Q) \Rrightarrow\\
\t1 \all s_1, s_2\mem P, t \leqslant c \cdot \all s_1'\cdot s_1 \sim s_2 \land s_1 \fun_t s_1' \implies \exists s_2'\cdot s_2 \fun_t s_2' \land s_1'\sim s_2'\]
where $t \leqslant c$ denotes that $t$ is a prefix of $c$,  $s_1\sim s_2$ denotes $s_1$ and $s_2$ agree on non-sensitive values, and $s_1\fun_t s_1'$ denotes $s_1'$ is reached from $s_1$ by instructions $t$. Note that since the programming language is deterministic, the above property implies that all states reached from $s_2$ by $t$ agree with the non-sensitive values of $s_1'$.

To support its use in a concurrent setting, $\wps$ also supports rely/guarantee reasoning \cite{jon83,xu97}. To detect additional vulnerabilities that arise due to a processor's memory model, it is paired with a notion of reordering interference freedom (rif) \cite{cou21,cou23}. These techniques (see \cite{cou24} for details) are independent of the details of the logic's rules and can be equally applied to the extensions to $\wps$ in this paper.

\section{Data Flow Spectre Variants}
\label{sec:data}

In addition to attacks related to speculation on control flow, such as Spectre-PHT of Section~\ref{sec:intro}, attacks have been identified based on speculation on data flow; specifically speculation on dependencies between stores and subsequent loads. The most well-known of these is Spectre variant 4 (also known as Spectre-STL)~\cite{can19}. This attack relies on a processor's memory disambiguator mispredicting that a load is independent of an earlier store, and hence executing the load with a stale value. 

\subsection{Spectre-STL}
\label{sec:ex_stl}

We illustrate Spectre-STL on Case 4 of the 13 litmus tests developed by Daniel et al.\ \cite{dan21} and available at \url{https://github.com/binsec/haunted_bench/blob/master/src/litmus-stl/programs/spectrev4.c}. The test is reexpressed in the language from Section~\ref{sec:wp}. In the code below, $idx$ is an input provided by the user who may be an attacker, $secretarray$ is a publicly inaccessible array containing sensitive data and has length $array\_size$, and $publicarray2$ is a publicly accessible array which has length 512*256 (512 is the cache line size in bits, and 256 the number of integers representable using 8 bits). 

\[r0 := idx;\\
r1 := array\_size;\\
r0 := r0 ~\&~ (r1 -1);\\
	secretarray[r0] := 0;\t1	\mbox{// This store may be bypassed}\\
	r1 := secretarray[r0];\\
	r2 := publicarray2[r1*512];\\
\]

The code begins by calculating the bitwise AND of $idx$ and $array\_size-1$ to obtain a valid index of $secretarray$. This avoids an array bounds bypass as in the Spectre-PHT attack. The value at the calculated index is set to 0, a non-sensitive value. This value is then read and used to read a value from $publicarray2$. The multiplication by 512 in the final step allows the value read from $secretarray$ to be deduced via a subsequent timing attack (by detecting the cache line affected by the read of $publicarray2$). 

This code is secure provided the value used to read $publicarray2$ is the non-sensitive value 0. However, if it is run and the memory disambiguator mispredicts that the load of $secretarray[r0]$ is independent of the prior store then the load can be executed first. In this case, a sensitive value will be used in the read from $publicarray2$. To prevent bypassing the store in this way, a typical mitigation is to insert a fence instruction after the store to $secretarray$ \cite{pon22}.

\subsection{Spectre-PSF}
\label{sec:ex_psf}

Spectre-PSF is a variant of Spectre-STL where, rather than mispredicting that a dependency does not exist between a load and earlier store, the memory disambiguator mispredicts that a dependency {\em does\/} exist \cite{cau20,pon22}. This behaviour has been confirmed as being possible on the AMD Zen 3 processor. We illustrate Spectre-PSF via an exploitable gadget from \cite{pon22} (based on example code from AMD). The gadget is reexpressed in the language of Section~\ref{sec:wp}. In the code below, $idx$ is an input provided by the user, $A$ is a public array of size 16, $C$ is a public array of length $C\_size$=2 initialised to [0,0], and $B$ is a public array of size 512*256. 

\[r0 := idx;\\
r1 := C\_size;\\
\ifc{(r0 < r1)}\,\{\\
\t1 C[0] := 64;\\
\t1 r1 := C[r0];\t1 	\mbox{// Value 64 may be forwarded to r1}\\
\t1 r1 := A[r1*r0];\\
\t1 r2 := B[r1*512];\\
\}
\]

Ignoring speculation on the branch, the code is secure provided that the value loaded from $C[r0]$ is 64 only when $idx$ (and hence $r0$) is 0: the value loaded from $A$ will be the publicly accessible value at index 0 when $idx$ is either 0 or 1. However, if the processor mispredicts a dependency between the store to $C[0]$ and the load from $C[r0]$ when $idx$ is 1 then the value 64 can be (incorrectly) forwarded to the load. That is, $r1$ will be set to 64 and subsequently value $A[64]$ will be used in the index of $B$ in the final load. This access of $A$ will be out of bounds and hence to potentially sensitive data. Again, a fence after the store can be used to mitigate the vulnerability.

\section{Detecting Spectre-STL}
\label{sec:stl}

The $\wps$ logic in Section~\ref{sec:wp} assumes that speculation starts only at branching points (of if statements or while loops). To detect the data flow variants of Spectre, we need to also allow speculation to start at stores. For Spectre-STL, when a store is reached during execution, we can begin speculating that it is not required for the following code, and hence can be bypassed (the store executing later after the following code). 

Given the code $s; c_1; c_2$ where $s$ is a store and $c_1$ and $c_2$ are sequences of instructions, when the code of $c_1$ is not dependent on $s$, speculation over $c_1$ will lead to the execution $c_1; s; c_2$, where $s$ has effectively been reordered after the instructions in $c_1$. When one or more instructions in $c_1$ are dependent on $s$, speculating over $c_1$ will lead to the execution ${\bf spec}(c_1); s; c_1; c_2$, where ${\bf spec}(c_1)$ includes rolling back the speculation's effects and hence has no affect on the program, but may alter the processor's microarchitecture. 

In practice, the number of instructions in $c_1$ is limited by the processor's {\em speculation window\/}, i.e., the upper bound on the number of instructions that can execute speculatively. This bound will depend on the microarchitectural components involved in the speculation. For Spectre-STL, it will depend on the size of the store buffer where bypassed store instructions wait to be executed, i.e., committed to memory. The size of this buffer can be up to 106 stores\footnote{\url{https://www.anandtech.com/show/16226/apple-silicon-m1-a14-deep-dive/2}} and hence, in general, beyond the size of the single procedures we are targeting in our work.   
Hence, as in $\wps$ we assume speculation can continue to the end of our code and do not explicitly model a speculation window. This results in a logic that is sound (as we check vulnerabilities within {\em any\/} sized speculation window), but can lead to false positives in cases where the actual speculation window is shorter than the code remaining to be executed.



To extend $\wps$ to detect Spectre-STL vulnerabilities, we modify the store rule as follows.

\begin{enumerate}[(i)]
\item The speculative postcondition $Q_s$ is added to the non-speculative precondition. By transferring the speculative {\em post\/}condition, we effectively ignore the store, reflecting that it does not occur as part of the speculation. As in the if statement rule, all ghost variables $\df{y}$ are replaced by false (to leave just the predicates in terms of the base variables) and each base variable $y^\base$ is replaced by $y$ (to identify these variables with variables in the non-speculative predicate). 
\item The speculative postcondition is also added to the speculative precondition. This reflects the case where the speculation on the store occurs in the context of an ongoing speculative execution. In this case, the store (being bypassed) will have no effect on the ongoing execution. For example, the rule will not cause a proof obligation $\Gamma_x = \bot$ to be resolved by a store $x:=0$ (where the literal 0 is a non-sensitive value).
\end{enumerate}
The resulting rule is formalised below (where the additions to the original store rule from Section~\ref{sec:wp}, corresponding to (i) and (ii) above, are underlined).

\[\wpstl(x := e, \lseq Q_s, Q\rseq) =
\lseq \M \underline{Q_s} \land Q_s\rename{x, \Gamma_x, \df{x}}{e, \Gamma_E(e), true},\\
Q\rename{x, \Gamma_x}{e, \Gamma_E(e)} \land \Gamma_E(e) \sqsubseteq \calL(x) \land & (1)\\
(\all y \cdot \Gamma_y \sqcap \calL(y) \sqsubseteq \calL(y)\rename{x}{e})\land\\
\underline{Q_s\rename{var^\base, d_1, ..., d_n}{var, false, ..., false}}\,\rseq\O\]
where $d_1,...,d_n$ is the list of ghost variables of the form $\df{y}$.

To illustrate the utility of this rule, we apply it (along with other rules of $\wps$) to the litmus test from Section~\ref{sec:ex_stl} in Figure~\ref{fig:ex_stl}, and to the same litmus test with a fence inserted to prevent speculation in Figure~\ref{fig:ex_stl_fence}. In both cases, a leak instruction is added before the access to $publicarray2$.

\begin{figure}[t]
\[\lseq \M (\df{idx} \implies\\
~(secretarray[idx \& (array\_size -1)]_{def} \implies \Gamma_{secretarray[idx \& (array\_size -1)]} = \bot) \land\\
 ~(\neg secretarray[idx \& (array\_size -1)]_{def} \implies \Gamma_{secretarray[idx \& (array\_size -1)]^\base} = \bot))\land\\
 (\neg\df{idx} \implies\\
 ~(secretarray[idx^\base \& (array\_size -1)]_{def} \implies \Gamma_{secretarray[idx^\base \& (array\_size -1)]} = \bot) \land\\
 ~(\neg secretarray[idx^\base \& (array\_size -1)]_{def} \implies \Gamma_{secretarray[idx^\base \& (array\_size -1)]^\base} = \bot)),\\
  \Gamma_{secretarray[idx \& (array\_size -1)]} = \bot\O\rseq\\
\codebox{r0 := idx;}\\
\lseq \M (secretarray[r0 \& (array\_size -1)]_{def} \implies \Gamma_{secretarray[r0\& (array\_size -1)]} = \bot) \land\\
 (\neg secretarray[r0 \& (array\_size -1)]_{def} \implies \Gamma_{secretarray[r0\& (array\_size -1)]^\base} = \bot),\\
  \Gamma_{secretarray[r0 \& (array\_size -1)]} = \bot\rseq\\
\codebox{r1 := array\_size;} \t1 // \mbox{ $array\_size=array\_size^\base$ since $array\_size$ is a constant}\\
\lseq \M (secretarray[r0\& (r1 -1)]_{def} \implies \Gamma_{secretarray[r0\& (r1 -1)]} = \bot) \land\\
 (\neg secretarray[r0\& (r1 -1)]_{def} \implies \Gamma_{secretarray[r0\& (r1 -1)]^\base} = \bot),\\
  \Gamma_{secretarray[r0 \& (r1 -1)]} = \bot\O\rseq\\
\codebox{r0 := r0 ~\&~ (r1 -1);}\\
\lseq \M(secretarray[r0]_{def} \implies \Gamma_{secretarray[r0]} = \bot) \land\\
 (\neg secretarray[r0]_{def} \implies \Gamma_{secretarray[r0]^\base} = \bot),\, \Gamma_{secretarray[r0]} = \bot\O\rseq\\
	\codebox{secretarray[r0] := 0 ;\t1	\mbox{// This store may be bypassed}}\\
	\lseq \M(secretarray[r0]_{def} \implies \Gamma_{secretarray[r0]} = \bot) \land\\
	(\neg secretarray[r0]_{def} \implies \Gamma_{secretarray[r0]^\base} = \bot),\, \Gamma_{secretarray[r0]} = \bot\O\rseq\\
	\codebox{r1 := secretarray[r0];}\\
	\lseq \Gamma_{r1} = \bot, \Gamma_{r1} = \bot\rseq\\
	\codebox{\leak r1;}\\
	\lseq true, true\rseq\\
	\codebox{r2 := publicarray2[r1*512];}\\
	\lseq true, true\rseq
\]
\vspace{-6mm}
\caption{Spectre-STL litmus test (code highlighted in \codebox{gray}).}
\vspace{-4mm}
\label{fig:ex_stl}
\end{figure}

The introduced leak instruction adds proof obligations in both the speculative and non-speculative states that $\Gamma_{r1}$ is $\bot$. Preceding backwards through the proof of Figure~\ref{fig:ex_stl}, these obligations are transformed by the load to $r1$ to conditions on $secretarray[r0]$; in the speculative case this condition is dependent on whether $secretarray[r0]$ is defined during the speculation. 

The interesting step is the store to $secretarray[0]$. Since the value stored is non-sensitive, the proof obligation is satisfied in the non-speculative case (assuming $secretarray[0]$ is not used in the security classification $\calL$ of another variable). Hence, the non-speculative precondition of the store includes only the transferred condition from the speculative postcondition, i.e., $\Gamma_{secretarray[r0]}=\bot$. The speculative precondition is equivalent to the speculative postcondition: the second conjunct of the precondition in rule~(1) evaluates to true when $secretarray[r0]$ is defined and $secretarray[r0]$ is non-sensitive. 

Proceding further backwards through the proof, the index used to access $secretarray$ is replaced with $idx\,\&\,(array\_size-1)$. Thus, the final non-speculative precondition is $\Gamma_{secretarray[idx\,\&\,(array\_size-1)]} = \bot$, indicating that the code is secure provided that this condition holds initially. This is more precise than a simple syntactic analysis which identifies the gadget, but does not define the conditions under which it can be successfully exploited. 

\begin{figure}[!t]
\[\lseq true,\, true\rseq\\
\codebox{r0 := idx;}\\
\lseq true,true\rseq\\
\codebox{r1 := array\_size;}\\
\lseq true, true\rseq\\
\codebox{r0 := r0 ~\&~ (r1 -1);}\\
\lseq true, true\rseq\\
	\codebox{secretarray[r0] := 0; \t1	\mbox{// This store may no longer be bypassed}}\\
	\lseq true, \Gamma_{secretarray[r0]} = \bot\rseq\\
	\codebox{\fence;}\\
	\lseq \M (secretarray[r0]_{def} \implies \Gamma_{secretarray[r0]} = \bot) \land\\
	 (\neg secretarray[r0]_{def} \implies \Gamma_{secretarray^\base[r0]} = \bot),\, \Gamma_{secretarray[r0]} = \bot\O\rseq\\
	\codebox{r1 := secretarray[r0];}\\
	\lseq \Gamma_{r1} = \bot, \Gamma_{r1} = \bot\rseq\\
	\codebox{\leak r1;}\\
	\lseq true, true\rseq\\
	\codebox{r2 := publicarray2[r1*512];}\\
	\lseq true, true\rseq
\]
\vspace{-6mm}
\caption{Spectre-STL litmus test with fence mitigation applied.}
\vspace{-4mm}
\label{fig:ex_stl_fence}
\end{figure}

The proof in Figure~\ref{fig:ex_stl_fence} is identical before the fence instruction is reached (i.e., below the fence instruction). At this point, the speculative predicate becomes true and hence no condition is transferred to the non-speculative precondition at the store instruction. The result is that the final non-speculative precondition is true, indicating that the code is always secure. 

To further validate our rule, we applied it (along with other required rules from $\wps$) to the remaining 12 litmus tests of Daniel et al.\ \cite{dan21} (see Appendix~A) and for each of the 9 litmus tests with a vulnerability, we applied it to a version of the litmus test with a fence added as a mitigation. All vulnerabilities were detected and all tests with mitigations showed the vulnerability could no longer occur. However, there is one test where we detect a vulnerability and Daniel et al.\ do not. This test, Case 9, is the same as Case 4 but includes a loop after the store which is intended to fill the reorder buffer\footnote{The reorder buffer contains {\em all\/} speculated instructions and provides an upper limit on the number of instructions that can be speculatively executed.}, forcing the store to be evaluated and take effect in memory before the load from $publicarray2$. Since our logic supports detection of Spectre-PHT (as well as Spectre-STL), it allows the loop to speculatively exit early. In general, our logic detects multiple variants of Spectre including, as in this case, vulnerabilities that arise due to their combination.


\section{Detecting Spectre-PSF}
\label{sec:psf}

Store forwarding refers to using the value of a store instruction in a subsequent load instruction before the store has taken effect in memory. This can be done safely when the store and load are to the same address. On some processors, store forwarding can be done speculatively based on a prediction that a store and subsequent load are to the same address. This leads to the Spectre-PSF vulnerability described in Section~\ref{sec:ex_psf}.

Abstracting from how the prediction is made, our rule reflects that the value of a store instruction, can be used speculatively in {\em any\/} subsequent load. When there is no leak or a given load does not cause a leak, the misprediction is benign and does not manifest in our reasoning. When the load does cause a leak, the variable associated with the load will appear in the postcondition of the store. For each subset of such variables, we replicate the speculative proof obligation with the variables replaced by the value of the store. This captures all possible predictions including those in which the value of the store is forwarded to more than one subsequent load. These additional proof obligations are also transferred to the non-speculative precondition of the store, reflecting that speculation may have begun at the store. The rule is formalised below (with the additions to the Spectre-STL store rule from Section~\ref{sec:stl} underlined).

\[\wppsf(x := e, \lseq Q_s, Q\rseq) =\\
\t1 \lseq \M \underline{\forall \{y_1,...,y_m\}\subseteq vars(Q_s)\cdot}\\
\t1 (Q_s \land Q_s\rename{x, \Gamma_x, \df{x}}{e, \Gamma_E(e), true})\underline{\rename{y_1,...,y_m}{e,...,e}},\\
Q\rename{x, \Gamma_x}{e, \Gamma_E(e)} \land \Gamma_E(e) \sqsubseteq \calL(x) \land & (2)\\
(\all y \cdot \Gamma_y \sqcap \calL(y) \sqsubseteq \calL(y)\rename{x}{e})\land\\
\underline{\forall \{y_1,...,y_m\} \subseteq vars(Q_s) \cdot}\\
\t1 Q_s\rename{var^\base, d_1, ..., d_n}{var, false, ..., false}\underline{\rename{y_1,...,y_m}{e,...,e}}\,\rseq\O\]
where $vars(\calL(x))$ denotes the list of variables occurring free in $\calL(x)$,
and $d_1,...,d_n$ is the list of ghost variables of the form $\df{y}$. Note that when the set $\{y_1,...,y_m\}$ is the empty set, the predicate in both the speculative and non-speculative preconditions are equivalent to those of the STL rule. Hence, this rule will detect vulnerabilities to both Spectre-STL and Spectre-PSF.

\begin{figure}[!t]
\[
\lseq ..., idx < C\_size \implies \M \Gamma_{A[C[idx]*idx]}=\bot \land \Gamma_{A[64*idx]} = \bot \land \Gamma_{C[idx]}=\bot \land\\
 \Gamma_{A[C[0\mapsto 64][idx]*idx]} = \bot)\O\land
 (idx \geq C\_size \implies ...)\O\rseq\\
\codebox{r0 := idx;}\\
\lseq ..., \M \Gamma_{r0} = \bot \land 
(r0 < C\_size \implies \M \Gamma_{A[C[r0]*r0]}=\bot \land \Gamma_{A[64*r0]} = \bot \land \Gamma_{C[r0]}=\bot \land\\
 \Gamma_{A[C[0\mapsto 64][r0]*r0]} = \bot)\O\land
 (r0 \geq C\_size \implies ...)\O\rseq\\
\codebox{r1 := C\_size;}\\
\lseq ..., \M \Gamma_{r0} = \bot \land \Gamma_{r1}=\bot \land\\
(r0 < r1 \implies \M \Gamma_{A[C[r0]*r0]}=\bot \land \Gamma_{A[64*r0]} = \bot \land \Gamma_{C[r0]}=\bot \land\\
 \Gamma_{A[C[0\mapsto 64][r0]*r0]} = \bot)\O\land
 (r0 \geq r1 \implies ...)\O\rseq\\
\codebox{\ifc{(r0 < r1)}\{}\\
\t1 \lseq ..., \M \Gamma_{A[C[r0]*r0]}=\bot \land \Gamma_{A[64*r0]} = \bot \land \Gamma_{r0}=\bot \land \Gamma_{C[r0]}=\bot \land\\
 \Gamma_{A[C[0\mapsto 64][r0]*r0]} = \bot\rseq\O\\
\t1 \codebox{C[0] := 64; \t1 //	\mbox{ Value 64 may be forwarded to r1}}\\
\t1 \lseq  \mbox{as below}\rseq\\
\t1 \codebox{\leak r0;}\\
\t1 \lseq \M(A[C[r0]*r0]_{def} \land C[r0]_{def} \implies \Gamma_{A[C[r0]*r0]} = \bot \land \Gamma_{r0}=\bot \land \Gamma_{C[r0]}=\bot) \land\\
 (A[C[r0]*r0]_{def} \land \neg C[r0]_{def} \implies \Gamma_{A[C[r0]^\base*r0]} = \bot \land \Gamma_{r0}=\bot \land \Gamma_{C[r0]^\base}=\bot) \land\\
(\neg A[C[r0]*r0]_{def} \land C[r0]_{def} \implies \Gamma_{A[C[r0]*r0]^\base} = \bot \land \Gamma_{r0}=\bot \land \Gamma_{C[r0]}=\bot) \land\\
(\neg A[C[r0]*r0]_{def} \land \neg C[r0]_{def} \implies \Gamma_{A[C[r0]^\base*r0]^\base} = \bot \land \Gamma_{r0}=\bot \land \Gamma_{C[r0]^\base}=\bot),\O\\
\t2 \Gamma_{r0}=\bot \land \Gamma_{C[r0]}=\bot \land \Gamma_{A[C[r0]*r0]} = \bot\rseq\\
\t1 \codebox{r1 := C[r0];}\\
\t1 \lseq \M \Gamma_{r0}=\bot \land \Gamma_{r1}=\bot \land (A[r1*r0]_{def} \implies \Gamma_{A[r1*r0]} = \bot) \land\\
(\neg A[r1*r0]_{def} \implies \Gamma_{A[r1*r0]^\base} = \bot),\, \Gamma_{r0}=\bot \land \Gamma_{r1}=\bot \land \Gamma_{A[r1*r0]} = \bot\O\rseq\\
\t1 \codebox{\leak r1*r0;}\\
\t1 \lseq (A[r1*r0]_{def} \implies \Gamma_{A[r1*r0]} = \bot) \land (\neg A[r1*r0]_{def} \implies \Gamma_{A[r1*r0]^\base} = \bot),\\
\t2 \Gamma_{A[r1*r0]} = \bot\rseq\\
\t1 \codebox{r1 := A[r1*r0];}\\
\t1 \lseq \Gamma_{r1} = \bot, \Gamma_{r1} = \bot\rseq\\
\t1 \codebox{\leak r1;}\\
\t1 \lseq true, true\rseq\\
\t1 \codebox{r1 := B[r1];}\\
\t1 \lseq true, true\rseq\\
\codebox{\}}\\
\lseq true, true\rseq
\]
\vspace{-6mm}
\caption{Spectre-PSF litmus test}
\label{fig:ex_psf}
\vspace{-4mm}
\end{figure}

To illustrate rule~(2), we apply it to the litmus test from Section~\ref{sec:ex_psf} in Figure~\ref{fig:ex_psf}. Each load of an array value ($B[r1]$, $A[r1*r0]$ and $C[r0]$) introduces a potential leak. Note that the leak due to the load of $C[r0]$ does not change the state tuple $\lseq Q_s,Q\rseq$ since both the non-speculative and speculative predicates already imply $\Gamma_{r0} = \bot$. For the non-speculative precondition of the store $C[0]:=64$, the postcondition $\Gamma_{r0}=\bot$ is unchanged, the postcondition $\Gamma_{C[r0]}=\bot$ is transformed to true, and the postcondition $\Gamma_{A[C[r0]*r0]}=\bot$ is transformed to $\Gamma_{A[C[0\mapsto 64][r0]*r0]}=\bot$ where $C[0\mapsto 64]$ is array $C$ with element 0 equal to 64. 

In addition, for each subset of global variables in the non-speculative postcondition, we need to transfer the required predicate to the non-speculative precondition. There are two global variables, $A[C[r0]*r0]$ and $C[r0]$, and hence four subsets including the empty set. The speculative postcondition with variables of the form $\df{y}$ set to false, and variables of the form $y^\base$ replaced by $y$ is $\Gamma_{A[C[r0]*r0]} = \bot \land \Gamma_{r0}=\bot \land \Gamma_{C[r0]}=\bot$. Hence, the required predicates for each subset of global variables are as follows.

\begin{itemize}
\item For the empty set $\{\}$, we have $\Gamma_{A[C[r0]*r0]} = \bot \land \Gamma_{r0}=\bot \land \Gamma_{C[r0]}=\bot$.
\item For $\{A[C[r0]*r0]\}$, we have $\Gamma_{r0}=\bot \land \Gamma_{C[r0]}=\bot$ since $\Gamma_{A[C[r0]*r0]} = \bot$ is true when $A[C[r0]*r0]$ is 64.
\item For $\{C[r0]\}$, we have $\Gamma_{A[64*r0]} = \bot \land \Gamma_{r0}=\bot$ since $\Gamma_{C[r0]} = \bot$ is true when $C[r0]$ is 64.
\item For $\{A[C[r0]*r0], C[r0]\}$, we have $\Gamma_{r0}=\bot$.
\end{itemize}
Conjoining the four predicates above gives us the condition required for the leaks not to be exploitable via either Spectre-STL (the empty-set case) or Spectre-PSF: $\Gamma_{A[C[r0]*r0]}=\bot \land \Gamma_{A[64*r0]} = \bot \land \Gamma_{r0}=\bot \land \Gamma_{C[r0]}=\bot$. 

The overall non-speculative precondition for the program is derived under the assumptions that $C\_size$ and the input $idx$ are non-sensitive, and that $idx \geq 0$ and all elements of arrays $A$ and $C$ are non-sensitive. To keep the presentation simple, we elide the speculative precondition above the store $C[0]:=64$ and the non-speculative proof obligation due to Spectre-PHT, i.e., the non-speculative proof obligation when $idx \geq C\_size$.

When $idx=0$, the precondition simplifies to true since each array index evaluates to 0 which is in the range of the respective arrays (recall from Section~\ref{sec:ex_psf} that $C$ is of size 2 and $A$ of size 16). When $idx=1$, $A[64*idx]$ accesses a memory location beyond the end of array $A$ and hence data which is potentially sensitive. Hence, the code is not provably secure: the Spectre-PSF vulnerability discussed in Section~\ref{sec:ex_psf} is detected. 

Adding a fence instruction after the store will prevent speculative store forwarding. This situation is also correctly evaluated by our logic. The fence's speculative precondition is true and hence no proof obligations are transferred to the non-speculative precondition of the store. This results in the precondition of the program when $idx$ is 0 or 1 evaluating to true. 

The above litmus test (in both fenced and unfenced form) constitutes the only litmus test for Spectre-PSF in the literature. Other approaches for detecting Spectre-PSF are based on an explicit semantics of the microarchitectural features that give rise to the vulnerability \cite{gua20} or, like us, rely on this single litmus test for validation \cite{pon22}.

\section{Related Work}
\label{sec:related}

Cauligi et al.\ \cite{cau22} provide a detailed comparison of 24 formal semantics and tools for detecting Spectre vulnerabilities. While all approaches support Spectre-PHT, only 5 out of 24 \cite{cau20,gua20,dan21,bar21,pon22} support Spectre-STL (and only 2 of these \cite{gua20,pon22} have support for Spectre-PSF). Three of the five are based on explicit models of a processor's microarchitecture \cite{cau20,gua20,dan21} and two on more abstract semantics \cite{bar21,pon22}.

Of the former approaches, Cauligi et al. \cite{cau20} and Guanciale et al.\ \cite{gua20} model program instructions by translation to sequences of fetch, execute and commit microinstructions. Additional state information and associated microinstructions provide the prediction and rollback facilities required to model speculative execution. This level of detail has the potential to detect more vulnerabilities than abstract approaches, and in fact Guanciale et al.\ \cite{gua20} independently discover Spectre-PSF (which they call Spectre-STL-D). However, such detailed models also add complexity to analysis. 

Cauligi et al.'s approach \cite{cau20} is supported by symbolic execution as is the approach of Daniel et al.\ \cite{dan21}. The latter work addresses scalability issues inherent with symbolic execution by removing redundant execution paths, and representing aspects of the microarchitectural execution symbolically rather than explicitly. These optimisations are validated using a set of litmus tests including those for Spectre-STL that we adopt in this paper. Later work by the authors~\cite{dan23} looks at modelling and implementing a hardware taint-tracking mechanism to mitigate vulnerabilities to Spectre, including Spectre-STL. 

Fabian et al.\ \cite{fab22} (not included in the above comparison) also employ symbolic execution for detecting Spectre vulnerabilities, including Spectre-STL. They define a framework for composing semantics of different variants of Spectre allowing to detect leaks due to a combination of, for example, Spectre-PHT and Spectre-STL. Our approach also allows the detection of such vulnerabilities as the proof obligations for each of the different Spectre variants is checked. We have confirmed this by applying the approach to Listing~1 of \cite{fab22} (see Appendix~B).

Barthe et al. \cite{bar21} provide a higher-level semantics of speculative execution for a simple while language (similar to the language in this paper). Rather than modelling speculation via microinstructions, the semantics includes high-level directives which, for example, force a particular branch to be taken, or indicate which store is to be used by a load. The approach is implemented in the Jasmin verification framework \cite{alm17,alm20}.

Ponce de Le\'on and Kinder \cite{pon22} provide an axiomatic semantics for speculative execution (based on the work of Alglave et al.\ \cite{alg14}) which is significantly less complex than the operational semantics of other approaches. The semantics defines which executions are valid via constraints on various relations between loads and stores in a program. Their approach is validated for Spectre-STL and Spectre-PSF using the same litmus tests as in this paper, and supported by bounded model checking.

Our work differs from the existing approaches by having its basis in weakest precondition (wp) reasoning. This opens the opportunity to adapt existing program analysis tools such as Boogie \cite{bar05a} or Why3 \cite{fil13} which automate such reasoning (see \cite{smi23} for work in this direction). Such tooling requires the user to provide annotations, particularly loop invariants, to programs but is able to handle greater nondeterminism than symbolic execution or model checking where nondeterminism can adversely affect scalability. 

Our work is also based on an approach \cite{cou24} which can be combined with rely/guarantee reasoning for analysis of concurrent programs \cite{jon83,xu97} and the proof technique, reordering interference freedom (rif), for taking into account processor weak memory models \cite{cou21}. Its underlying logic can also be extended to support controlled release of sensitive information via declassification \cite{smi22}.
 
\section{Conclusion}
\label{sec:con}

This paper has presented a weakest precondition-based approach for detecting vulnerabilities to the major data flow variants of Spectre, Spectre-STL and Spectre-PSF. The approach extends an existing approach for Spectre-PHT and can detect vulnerabilities to all three attacks including when the attacks occur in combination. The approach has been validated with a set of litmus test used to validate related approaches and tools in the literature. A deeper evaluation of the approach, including its use on concurrent programs, requires automated tool support which is left to future work. Since it is based on weakest precondition reasoning, such support can be built on an existing auto-active program analyser such as Boogie or Why3.


\medskip

\noindent{\bf Acknowledgements} Thanks to Kirsten Winter, Robert Colvin and Mark Beaumont for feedback on this paper.

\bibliographystyle{splncs04}
\bibliography{references}


\newpage
\appendix


\section{Spectre-STL Litmus tests}

Below we apply our logic to the 13 litmus tests for Spectre-STL developed by Daniel et al.\ \cite{dan21} and available at \url{https://github.com/binsec/haunted_bench/blob/master/src/litmus-stl/programs/spectrev4.c}. The tests are reexpressed in the language from Section~\ref{sec:wp}. In the tests, $idx$ is an input provided by the user who may be an attacker. This value is not sensitive, i.e., $\calL(idx) =\bot$. $secretarray$ is a publicly inaccessible array which may contain sensitive data and has length $array\_size=16$. This value is not sensitive, i.e., $\calL(array\_size) = \bot$. $publicarray$ is a publicly accessible array which has length $array\_size$. For all indices $i$ of $publicarray$, $\calL(publicarray[i])=\bot$. $publicarray2$ is a publicly accessible array which has length 512*256 (512 is the cache line size in bits, and 256 the number of integers representable using 8 bits). For all indices $i$ of $publicarray2$, $\calL(publicarray2[i])=\bot$. Hence, the only sensitive data is in $secretarray$ or parts of memory outside of the defined arrays and variables. 

The speculative precondition is elided whenever there are no points where speculation can start (stores or branches) earlier in the code.

\subsection*{Case 1} $data$, $data\_slowptr$ and $data\_slowslowptr$ are local (pointer) variables and can point to sensitive data. To express this test in our simple programming language, all expressions involving referencing (\&) and dereferencing (*) of pointers have been resolved. 

The code is insecure since the non-speculative precondition requires an element of $secretarray$ to have security level $\bot$. 

\[\lseq   ...,\,
 \Gamma_{secretarray[idx \& (array\_size -1)]} = \bot\rseq\\
	r0 := idx;\\
 \lseq  ...,
\, \Gamma_{secretarray[r0 \& (array\_size -1)]} = \bot\rseq\\
	r1 := array\_size; 
	\\
\lseq  ..., 
 \, \Gamma_{secretarray[r0 \& (r1 -1)]} = \bot\rseq\\
	r0 := r0 ~\&~ (r1 -1);\\
\lseq  ..., 
 \, \Gamma_{secretarray[r0]} = \bot\rseq\\	
	r1 := secretarray;\\
\lseq ..., \, 
 \Gamma_{secretarray[r0]} = \bot\rseq\\	
	data := r1;\\
\lseq \M (\neg secretarray[r0]_{def} \implies \Gamma_{secretarray[r0]^\base} = \bot)\land \Gamma_{secretarray[r0]} = \bot,\\
 \Gamma_{secretarray[r0]} = \bot\O\rseq\\	
	data\_slowptr := r1;\zbreak
\lseq \M (\neg secretarray[r0]_{def} \implies \Gamma_{secretarray[r0]^\base} = \bot)\land \Gamma_{secretarray[r0]} = \bot,\\
\Gamma_{secretarray[r0]} = \bot\O\rseq\\	
	data\_slowslowptr := r1;\\
\lseq \M (\neg secretarray[r0]_{def} \implies \Gamma_{secretarray[r0]^\base} = \bot)\land \Gamma_{secretarray[r0]} = \bot,\\
  \Gamma_{secretarray[r0]} = \bot\O\rseq\\
	secretarray[r0] := 0 \t1	\mbox{// This store may be bypassed}\\
\lseq \M(secretarray[r0]_{def} \implies \Gamma_{secretarray[r0]} = \bot) \land\\
(\neg secretarray[r0]_{def} \implies \Gamma_{secretarray[r0]^\base} = \bot),\, \Gamma_{secretarray[r0]} = \bot\O\rseq\\
	r1 := secretarray[r0];\\
\lseq \Gamma_{r1} = \bot, \Gamma_{r1} = \bot\rseq\\
	\leak r1;\\
\lseq true,\, true\rseq\\
	r2 := publicarray2[r1*512];\\
\lseq true,\, true\rseq
\]

\subsection*{Case 2}

The code is insecure since $idx$ may be greater than the length of $publicarray$. 

\[\lseq \M ...,
\Gamma_{publicarray[idx]} = \bot\O\rseq\\
	r0 := idx;\\
\lseq \M ...,\, 
 \Gamma_{publicarray[r0 \& (array\_size -1)]} = \bot \land \Gamma_{r0}=\bot \land \Gamma_{publicarray[idx]} = \bot\O\rseq\\
	r1 := array\_size; 
	\\
\lseq \M ...,\, 
 \Gamma_{publicarray[r0 \& (r1 -1)]} = \bot \land \Gamma_{r0}=\bot \land \Gamma_{r1}=\bot \land \Gamma_{publicarray[idx]} = \bot\O\rseq\\
	r0 := r0 ~\&~ (r1 -1);\\
\lseq \M ...,\, 
 \Gamma_{publicarray[r0]} = \bot \land \Gamma_{r0}=\bot\land \Gamma_{publicarray[idx]} = \bot\O\rseq\\
	idx := r0 \t1	\mbox{// This store may be bypassed}\\
\lseq \M(publicarray[idx]_{def} \implies \Gamma_{publicarray[idx]} = \bot) \land\\
(\neg publicarray[idx]_{def} \implies \Gamma_{publicarray[idx]^\base} = \bot),\, \Gamma_{publicarray[idx]} = \bot\O\rseq\\
	\leak idx; \t1 // \mbox{~note that $\Gamma_{idx}=\bot$ is true since $\calL(idx)=\bot$}\\
\lseq \M(publicarray[idx]_{def} \implies \Gamma_{publicarray[idx]} = \bot) \land\\
(\neg publicarray[idx]_{def} \implies \Gamma_{publicarray[idx]^\base} = \bot),\, \Gamma_{publicarray[idx]} = \bot\O\rseq\\
	r1 := publicarray[idx];\\
\lseq \Gamma_{r1} = \bot, \Gamma_{r1} = \bot\rseq\\
	\leak r1;\\
\lseq true,\, true\rseq\\
	r2 := publicarray2[r1*512];\\
\lseq true,\, true\rseq
\]

\subsection*{Case 3}

The code is secure since there is no store to bypass. 

\[\lseq ...,\, true\rseq\\
	r0 := idx;\zbreak
\lseq ... 
,\, \Gamma_{r0}=\bot\rseq\\
	r1 := array\_size; 
	\\
\lseq \M... 
,\,
\Gamma_{r0}=\bot \land\Gamma_{r1}=\bot \land\Gamma_{publicarray[r0 \& (r1 -1)]} = \bot\O\rseq\\
	r0 := r0 ~\&~ (r1 -1);\\
\lseq \M...,\, 
\Gamma_{r0}=\bot \land\Gamma_{publicarray[r0]} = \bot\O\rseq\\	
	\leak r0;\\
\lseq \M...,\, 
\Gamma_{publicarray[r0]} = \bot\O\rseq\\
	r1 := publicarray[r0];\\
\lseq ...,\, 
\Gamma_{r1} = \bot\rseq\\
	\leak r1;\\
\lseq ...,\, 
true\rseq\\
	r2 := publicarray2[r1*512];\\
\lseq true, true\rseq
\]

\subsection*{Case 4}

See Section~\ref{sec:stl}.

\subsection*{Case 5}

$case5\_ptr$ and $toleak$ are local variables. $case5\_ptr$ is initially set to $secretarray$. 

The code is insecure since it requires that elements of $secretarray$  (the initial value of $case5\_ptr$) have security level $\bot$.

\[
 \lseq \M ...,\, 
\Gamma_{toleak} = \bot \land \Gamma_{case5\_ptr[idx \& (array\_size -1)]} = \bot \rseq\O\\
	r0 := idx;\t1 
	\\
 \lseq \M ...,\, 
\Gamma_{toleak} = \bot \land \Gamma_{case5\_ptr[r0 \& (array\_size -1)]} = \bot \rseq\O\\
	r1 := array\_size; 
	\\
 \lseq \M ...,\, 
\Gamma_{toleak} = \bot \land \Gamma_{case5\_ptr[r0 \& (r1 -1)]} = \bot \land \Gamma_{publicarray[r0 \& (r1 -1)]}=\bot\rseq\O\\
	r0 := r0 ~\&~ (r1 -1);\\
\lseq \M...,\, 
\Gamma_{toleak} = \bot \land \Gamma_{case5\_ptr[r0]} = \bot \land \Gamma_{publicarray[r0]}=\bot\rseq\O\\
	r1 := publicarray;\\
\lseq \M ...,\, 
\Gamma_{toleak} = \bot \land \Gamma_{case5\_ptr[r0]} = \bot \land \Gamma_{r1[r0]}=\bot\rseq\O\\
	case5\_ptr := r1;\t1 \mbox{// This store may be bypassed}\\
\lseq  \M(\df{case5\_ptr[r0]} \implies \Gamma_{case5\_ptr[r0]} = \bot) \land (\neg\df{case5\_ptr[r0]} \implies \Gamma_{case5\_ptr[r0]^\base} = \bot)\land \\\Gamma_{toleak} = \bot, \,\Gamma_{case5\_ptr[r0]} = \bot\land \Gamma_{toleak} = \bot\rseq\O\\
	r1 = case5\_ptr[r0];\\
\lseq \Gamma_{toleak} = \bot \land \Gamma_{r1} = \bot, \Gamma_{r1} = \bot\land \Gamma_{toleak} = \bot\rseq\\
	toleak = r1;\\
\lseq \Gamma_{toleak} = \bot, \Gamma_{toleak} = \bot\rseq\\
	r1 := toleak;\t1 // \mbox{ if $\neg\df{toleak}$ then $toleak=toleak^\base$ since $toleak$ is local}\\
\lseq \Gamma_{r1} = \bot, \Gamma_{r1} = \bot\rseq\\
	\leak r1;\\
\lseq true, true\rseq\\
	r2 := publicarray2[r1*512];\\
\lseq true, true\rseq
\]

\subsection*{Case 6}

$case6\_idx$, $case6\_array$ and $toleak$ are local variables. $case6\_idx$ is initially set to 0, and $case6\_array$ is initially set to $[secretarray, publicarray]$.

The code is insecure since it requires elements of $secretarray$ (the initial value of $case6\_array[case6\_idx]$) to have security level $\bot$.

\[	
\lseq \M
 ..., \,
 \Gamma_{toleak} = \bot\land 
  \Gamma_{case6\_array[case6\_idx][idx \& (array\_size -1)]} = \bot\rseq\O\\
	r0 := idx;
	\\
\lseq \M ...,\, 
 \Gamma_{case6\_array[1][r0 \& (array\_size -1)]} = \bot\land \Gamma_{toleak} = \bot\land \Gamma_{case6\_idx} = \bot\land\\
  \Gamma_{r0 \& (array\_size -1)} = \bot \Gamma_{case6\_array[case6\_idx][r0 \& (array\_size -1)]} = \bot\rseq\O\\
	r1 := array\_size; 
	\\
\lseq \M ...\, 
 \Gamma_{case6\_array[1][r0 \& (r1 -1)]} = \bot\land \Gamma_{toleak} = \bot\land \Gamma_{case6\_idx} = \bot\land \\
 \Gamma_{r0 \& (r1 -1)} = \bot\land 
  \Gamma_{case6\_array[case6\_idx][r0 \& (r1 -1)]} = \bot\rseq\O\\
	r0 := r0 ~\&~ (r1 -1);\\
\lseq \M ...\, 
 \Gamma_{case6\_array[case6\_idx][r0]} = \bot\land \Gamma_{toleak} = \bot\land \Gamma_{case6\_idx} = \bot\land\\
  \Gamma_{r0} = \bot\rseq\O\\
	case6\_idx := 1;\t1 \mbox{// This store may be bypassed}\\
\lseq \M(\df{case6\_array[case6\_idx][r0]} \implies \Gamma_{case6\_array[case6\_idx][r0]} = \bot) \land\\
(\neg\df{case6\_array[case6\_idx][r0]} \implies \Gamma_{case6\_array[case6\_idx][r0]^\base} = \bot) \land\\
\Gamma_{toleak} = \bot \land \Gamma_{case6\_idx} = \bot\land \Gamma_{r0} = \bot,\\
 \Gamma_{case6\_array[case6\_idx][r0]} = \bot\land \Gamma_{toleak} = \bot\land \Gamma_{case6\_idx} = \bot\land \Gamma_{r0} = \bot\rseq\O\\
	r1 := case6\_idx;\\
\lseq \M(\df{case6\_array[r1][r0]} \implies \Gamma_{case6\_array[r1][r0]} = \bot) \land\\
(\neg\df{case6\_array[r1][r0]} \implies \Gamma_{case6\_array[r1][r0]^\base} = \bot) \land\\
\Gamma_{toleak} = \bot \land \Gamma_{r1} = \bot\land \Gamma_{r0} = \bot,\\
 \Gamma_{case6\_array[r1][r0]} = \bot\land \Gamma_{toleak} = \bot\land \Gamma_{r1} = \bot\land \Gamma_{r0} = \bot\rseq\O\\
	\leak r0;\\
\lseq \M(\df{case6\_array[r1][r0]} \implies \Gamma_{case6\_array[r1][r0]} = \bot) \land\\
(\neg\df{case6\_array[r1][r0]} \implies \Gamma_{case6\_array[r1][r0]^\base} = \bot) \land\\
\Gamma_{toleak} = \bot \land \Gamma_{r1} = \bot, \, \Gamma_{case6\_array[r1])[r0]} = \bot\land \Gamma_{toleak} = \bot\land \Gamma_{r1} = \bot\rseq\O\\
	\leak r1;\\
\lseq \M(\df{case6\_array[r1][r0]} \implies \Gamma_{case6\_array[r1][r0]} = \bot) \land\\
(\neg\df{case6\_array[r1][r0]} \implies \Gamma_{case6\_array[r1][r0]^\base} = \bot) \land\\
\Gamma_{toleak} = \bot, \Gamma_{case6\_array[r1][r0]} = \bot\land \Gamma_{toleak} = \bot\rseq\O\\
	r2 := (case6\_array[r1])[r0];\\
\lseq \Gamma_{toleak} = \bot \land\Gamma_{r2} = \bot, \,\Gamma_{r2} = \bot\land \Gamma_{toleak} = \bot\rseq\\
	toleak = r2;\\
\lseq \Gamma_{toleak} = \bot, \,\Gamma_{toleak} = \bot\rseq\\
	r1 := toleak;\t1 // \mbox{ if $\neg\df{toleak}$ then $toleak=toleak^\base$ since $toleak$ is local}\\
\lseq \Gamma_{r1} = \bot,\, \Gamma_{r1} = \bot\rseq\\
	\leak r1;\\
\lseq true,\, true\rseq\\
	r2 := publicarray2[r1*512];\\
\lseq true,\, true\rseq
\]

\subsection*{Case 7}

$case7\_mask$ a local variable which is initially set to $MAXINT$. 

The code is insecure since $idx~\&~MAXINT$ (where $MAXINT$ is the initial value of $case7\_mask$) may be greater than the length of $publicarray$. 

\[\lseq \M ...,\,  \Gamma_{toleak} = \bot\land 
\Gamma_{publicarray[idx\&case7\_mask]} = \bot
\O\\
	r0:= array\_size; \\
\lseq \M ...,\, 
\Gamma_{publicarray[idx\&(r0-1)]} = \bot\land \Gamma_{toleak} = \bot\land \Gamma_{idx}=\bot \land \Gamma_{r0}=\bot \land \\
\Gamma_{publicarray[idx\&case7\_mask]} = \bot\land
 \Gamma_{case7\_mask}=\bot\rseq\O\\
	r0 := r0-1;\\
\lseq \M ... 
\Gamma_{publicarray[idx\&r0]} = \bot\land \Gamma_{toleak} = \bot\land \Gamma_{idx}=\bot \land \Gamma_{r0}=\bot \land \\
\Gamma_{publicarray[idx\&case7\_mask]} = \bot\land
 \Gamma_{case7\_mask}=\bot\rseq\O\\
	case7\_mask := r0; \t1 // \mbox{This store may be bypassed}\\
\lseq \M(\df{publicarray[idx\&case7\_mask]} \implies \Gamma_{publicarray[idx\&case7\_mask]} = \bot) \land\\
(\neg\df{publicarray[idx\&case7\_mask]} \implies \Gamma_{publicarray[idx\&case7\_mask]^\base} = \bot)\land \\
\Gamma_{toleak} = \bot \land \Gamma_{idx}=\bot \land \Gamma_{case7\_mask}=\bot,\\
\Gamma_{publicarray[idx\&case7\_mask]} = \bot\land \Gamma_{toleak} = \bot\land \Gamma_{idx}=\bot \land \Gamma_{case7\_mask}=\bot\rseq\O\\
	r0 := idx;\t1 // \mbox{ if $\neg\df{idx}$ then $idx=idx^\base$ since $idx$ is local}\\
\lseq \M(\df{publicarray[r0\&case7\_mask]} \implies \Gamma_{publicarray[r0\&case7\_mask]} = \bot) \land\\
(\neg\df{publicarray[r0\&case7\_mask]} \implies \Gamma_{publicarray[r0\&case7\_mask]^\base} = \bot)\land \\
\Gamma_{toleak} = \bot \land \Gamma_{r0}=\bot \land \Gamma_{case7\_mask}=\bot,\\
\Gamma_{publicarray[r0\&case7\_mask]} = \bot\land \Gamma_{toleak} = \bot\land \Gamma_{r0}=\bot \land \Gamma_{case7\_mask}=\bot\rseq\O\\
	r1:= case7\_mask;\t1 // \mbox{ if $\neg\df{case7\_mask}$ then $case7\_mask=case7\_mask^\base$ since local}\\
\lseq \M(\df{publicarray[r0\&r1]} \implies \Gamma_{publicarray[r0\&r1]} = \bot) \land\\
(\neg\df{publicarray[r0\&r1]} \implies \Gamma_{publicarray[r0\&r1]^\base} = \bot)\land 
\Gamma_{toleak} = \bot \land \Gamma_{r0}=\bot \land \Gamma_{r1}=\bot,\\
\Gamma_{publicarray[r0\&r1]} = \bot\land \Gamma_{toleak} = \bot\land \Gamma_{r0}=\bot \land \Gamma_{r1}=\bot\rseq\O\\
	\leak r0~\&~r1;\\
\lseq \M(\df{publicarray[r0\&r1]} \implies \Gamma_{publicarray[r0\&r1]} = \bot) \land\\
(\neg\df{publicarray[r0\&r1]} \implies \Gamma_{publicarray[r0\&r1]^\base} = \bot)\land
\Gamma_{toleak} = \bot,\\
\Gamma_{publicarray[r0\&r1]} = \bot\land \Gamma_{toleak} = \bot\rseq\O\\
	r1 := publicarray[r0 ~\&~ r1];\\
\lseq \Gamma_{toleak} = \bot \land\Gamma_{r1} = \bot, \,\Gamma_{r1} = \bot\land \Gamma_{toleak} = \bot\rseq\\
	toleak := r1;\\
\lseq \Gamma_{toleak} = \bot, \,\Gamma_{toleak} = \bot\rseq\\
	r1 := toleak;\t1 // \mbox{ if $\neg\df{toleak}$ then $toleak=toleak^\base$ since $toleak$ is local}\\
\lseq \Gamma_{r1} = \bot,\, \Gamma_{r1} = \bot\rseq\\
	\leak r1;\\
\lseq true,\, true\rseq\\
	r2 := publicarray2[r1*512];\\
\lseq true,\, true\rseq
\]

\subsection*{Case 8}

$case8\_mult$ is a local variable which is initially set to 200.

The code is insecure since $idx*200$ (where 200 is the initial value of $case8\_mult$) may be greater than the length of $publicarray$. 

\[	
\lseq ...,\, 
\Gamma_{toleak} = \bot \land \Gamma_{publicarray[idx*case8\_mult]} = \bot\rseq\\
	case8\_mult := 0; \t1 // \mbox{This store may be bypassed}\\
\lseq \M\Gamma_{idx} = \bot \land \Gamma_{case8\_mult} = \bot\land\\
(\df{publicarray[idx*case8\_mult]} \implies \Gamma_{publicarray[idx*case8\_mult]} = \bot) \land\\
(\neg\df{publicarray[idx*case8\_mult]} \implies \Gamma_{publicarray[idx*case8\_mult]^\base} = \bot)\land
\Gamma_{toleak} = \bot,\\
\Gamma_{idx} = \bot \land \Gamma_{case8\_mult} = \bot\land \Gamma_{publicarray[r0*case8\_mult]} = \bot\land \Gamma_{toleak} = \bot\rseq\O\\
	r0:=idx;\t1 // \mbox{ if $\neg\df{idx}$ then $idx=idx^\base$ since $idx$ is local}\\
\lseq \M\Gamma_{r0} = \bot \land \Gamma_{case8\_mult} = \bot\land\\
(\df{publicarray[r0*case8\_mult]} \implies \Gamma_{publicarray[r0*case8\_mult]} = \bot) \land\\
(\neg\df{publicarray[r0*case8\_mult]} \implies \Gamma_{publicarray[r0*case8\_mult]^\base} = \bot)\land
\Gamma_{toleak} = \bot,\\
\Gamma_{r0} = \bot \land \Gamma_{case8\_mult} = \bot\land \Gamma_{publicarray[r0*case8\_mult]} = \bot\land \Gamma_{toleak} = \bot\rseq\O\\
	r1 := case8\_mult;\t1 // \mbox{ if $\neg\df{case8\_mult}$ then $case8\_mult=case8\_mult^\base$ since local}\\
\lseq \M\Gamma_{r0} = \bot \land \Gamma_{r1} = \bot\land (\df{publicarray[r0*r1]} \implies \Gamma_{publicarray[r0*r1]} = \bot) \land\\
(\neg\df{publicarray[r0*r1]} \implies \Gamma_{publicarray[r0*r1]^\base} = \bot)\land
\Gamma_{toleak} = \bot,\\
\Gamma_{r0} = \bot \land \Gamma_{r1} = \bot\land \Gamma_{publicarray[r0*r1]} = \bot\land \Gamma_{toleak} = \bot\rseq\O\\
	\leak r0*r1;\\
\lseq \M(\df{publicarray[r0*r1]} \implies \Gamma_{publicarray[r0*r1]} = \bot) \land\\
(\neg\df{publicarray[r0*r1]} \implies \Gamma_{publicarray[r0*r1]^\base} = \bot)\land
\Gamma_{toleak} = \bot,\\
\Gamma_{publicarray[r0*r1]} = \bot\land \Gamma_{toleak} = \bot\rseq\O\\
	r0 := publicarray[r0*r1];\\
\lseq \Gamma_{toleak} = \bot \land\Gamma_{r0} = \bot, \,\Gamma_{r0} = \bot\land \Gamma_{toleak} = \bot\rseq\\
	toleak := r0;\\
\lseq \Gamma_{toleak} = \bot, \,\Gamma_{toleak} = \bot\rseq\\
	r1 := toleak;\t1 // \mbox{ if $\neg\df{toleak}$ then $toleak=toleak^\base$ since $toleak$ is local}\\
\lseq \Gamma_{r1} = \bot,\, \Gamma_{r1} = \bot\rseq\\
	\leak r1;\\
\lseq true,\, true\rseq\\
	r2 := publicarray2[r1*512];\\
\lseq true,\, true\rseq
\]

\subsection*{Case 9}

The code is insecure since it requires elements of $secretarray$ to have security level $\bot$.

\[
\lseq ...,\, \M \Gamma_{secretarray[idx\&(array\_size-1)]} = \bot\O\rseq\\
	r0 := idx;\\
\lseq ...,\, \M \Gamma_{secretarray[r0\&(array\_size-1)]} = \bot\O\rseq\\
	r1 := array\_size; 
	\\
\lseq \M ...,\, 
\Gamma_{secretarray[r0\&(r1-1)]} = \bot)\O\rseq\\
	r0 := r0 ~\&~ (r1-1);\zbreak
\lseq \M ...,\, 
 \Gamma_{secretarray[r0]} = \bot\O\rseq\\
	secretarray[r0] := 0; \t1 // \mbox{This store may be bypassed}\\
\lseq \M(secretarray[r0]_{def} \implies \Gamma_{secretarray[r0]} = \bot) \land\\
(\neg secretarray[r0]_{def} \implies \Gamma_{secretarray[r0]^\base} = \bot),\, \Gamma_{secretarray[r0]} = \bot\O\rseq\\
	i := 0;\t1 // \mbox{ if $\neg\df{i}$ then $i=i^\base$ since $i$ is local}\\
\lseq \M(secretarray[r0]_{def} \implies \Gamma_{secretarray[r0]} = \bot) \land\\
(\neg secretarray[r0]_{def} \implies \Gamma_{secretarray[r0]^\base} = \bot),\, \Gamma_{i}=\bot \land \Gamma_{secretarray[r0]} = \bot\O\rseq\\
	while (i < 200) \{\\
\t1 \lseq \M(secretarray[r0]_{def} \implies \Gamma_{secretarray[r0]} = \bot) \land\\
(\neg secretarray[r0]_{def} \implies \Gamma_{secretarray[r0]^\base} = \bot),\, \Gamma_{secretarray[r0]} = \bot\O\rseq\\
	\t1 r1 := i;\t1 // \mbox{ if $\neg\df{i}$ then $i=i^\base$ since $i$ is local}\\
\t1 \lseq \M(secretarray[r0]_{def} \implies \Gamma_{secretarray[r0]} = \bot) \land\\
(\neg secretarray[r0]_{def} \implies \Gamma_{secretarray[r0]^\base} = \bot),\, \Gamma_{secretarray[r0]} = \bot\O\rseq\\
	\t1 i := r1 + 1;\\
\t1 \lseq \M(secretarray[r0]_{def} \implies \Gamma_{secretarray[r0]} = \bot) \land\\
(\neg secretarray[r0]_{def} \implies \Gamma_{secretarray[r0]^\base} = \bot),\, \Gamma_{secretarray[r0]} = \bot\O\rseq\\
\}\\
\lseq \M(secretarray[r0]_{def} \implies \Gamma_{secretarray[r0]} = \bot) \land\\
(\neg secretarray[r0]_{def} \implies \Gamma_{secretarray[r0]^\base} = \bot),\, \Gamma_{secretarray[r0]} = \bot\O\rseq\\
	r1:= secretarray[r0];\\
\lseq \Gamma_{r1} = \bot,\, \Gamma_{r1} = \bot\rseq\\
	\leak r1;\\
\lseq true,\, true\rseq\\
	r2 := publicarray2[r1*512];\\
\lseq true,\, true\rseq
\]

\subsection*{Case 10}

The remaining cases involve function calls. We model these by inlining the function code which is followed by a fence. The fence ensures that all stores in the function are executed before it returns. We use the suffix $\_ret$ to denote the return value of the function. 

The code is insecure since $fidx$ may be greater than the length of $publicarray$. 

\[	
\lseq ...,\,  \Gamma_{publicarray[fidx]} = \bot\rseq\\
	r0 := idx;\\
\lseq ...,\,  \Gamma_{r0 \& (array\_size-1)}=\bot \land \Gamma_{publicarray[r0 \& (array\_size-1)]} = \bot \land \Gamma_{publicarray[fidx]} = \bot\rseq\\
	r1 := array\_size; 
	\\
\lseq ...,\,  \Gamma_{r0 \& (r1-1)}=\bot \land \Gamma_{publicarray[r0 \& (r1-1)]} = \bot \land \Gamma_{publicarray[fidx]} = \bot\rseq\\
	r0\_ret := r0 ~\&~ (r1-1);\\
\lseq ...,\,  \Gamma_{r0\_ret}=\bot \land \Gamma_{publicarray[r0\_ret]} = \bot \land \Gamma_{publicarray[fidx]} = \bot\rseq\\
	\fence;\\
\lseq \M  ...,\,
 \Gamma_{r0\_ret}=\bot \land \Gamma_{publicarray[r0\_ret]} = \bot \land \Gamma_{publicarray[fidx]} = \bot\O\rseq\\
	fidx := r0\_ret; \t1 // \mbox{This store may be bypassed}\zbreak
\lseq \M  \Gamma_{fidx}=\bot \land (\df{fidx} \implies\\
\t1 (publicarray[fidx]_{def} \implies \Gamma_{publicarray[fidx]} = \bot) \land\\
\t1(\neg publicarray[fidx]_{def} \implies \Gamma_{publicarray[fidx]^\base} = \bot))\land\\
(\neg\df{fidx} \implies\\
\t1 (publicarray[fidx]_{def} \implies \Gamma_{publicarray[fidx^\base]} = \bot) \land\\
\t1(\neg publicarray[fidx]_{def} \implies \Gamma_{publicarray[fidx^\base]^\base} = \bot)),\\
 \Gamma_{fidx}=\bot \land \Gamma_{publicarray[fidx]} = \bot\O\rseq\\
	r0 := fidx;\t1 // \mbox{ if $\neg\df{fidx}$ then $fidx=fidx^\base$ since $fidx$ is local}\\
\lseq \M \Gamma_{r0}=\bot \land (publicarray[r0]_{def} \implies \Gamma_{publicarray[r0]} = \bot) \land\\
(\neg publicarray[r0]_{def} \implies \Gamma_{publicarray[r0]^\base} = \bot),\, \Gamma_{r0}=\bot \land \Gamma_{publicarray[r0]} = \bot\O\rseq\\
	\leak r0;\\
\lseq \M(publicarray[r0]_{def} \implies \Gamma_{publicarray[r0]} = \bot) \land\\
(\neg publicarray[r0]_{def} \implies \Gamma_{publicarray[r0]^\base} = \bot),\, \Gamma_{publicarray[r0]} = \bot\O\rseq\\
	r1:= publicarray[r0];\\
\lseq \Gamma_{r1} = \bot,\, \Gamma_{r1} = \bot\rseq\\
	\leak r1;\\
\lseq true,\, true\rseq\\
	r2 := publicarray2[r1*512];\\
\lseq true,\, true\rseq
\]

\subsection*{Case 11}

The code is insecure since it requires that $toleak$ is initially non-sensitive. 

\[\lseq ...,\, \Gamma_{toleak} = \bot\rseq\\	
	r0 := idx;\\
\lseq ...,\, \Gamma_{r0 \& (array\_size-1)} \land \Gamma_{toleak} = \bot\land \Gamma_{publicarray[r0 \& (array\_size-1)]} = \bot\rseq\\
	r1 := array\_size; \\
\lseq ...,\, \Gamma_{r0 \& (r1-1)} \land \Gamma_{toleak} = \bot\land \Gamma_{publicarray[r0 \& (r1-1)]} = \bot\rseq\\
	r0 := r0 ~\&~ (r1-1);\\
\lseq ...,\, \Gamma_{r0} \land \Gamma_{toleak} = \bot\land \Gamma_{publicarray[r0]} = \bot\rseq\\
	\leak r0;\\
\lseq ...,\, \Gamma_{toleak} = \bot\land \Gamma_{publicarray[r0]} = \bot\rseq\\
	r1 := publicarray[r0];\\
\lseq ...,\, \Gamma_{toleak} = \bot\land \Gamma_{r1} = \bot\rseq\\
	toleak\_ret := r1;\\
\lseq true,\, \Gamma_{toleak} = \bot\land \Gamma_{toleak\_ret} = \bot\rseq\\
	\fence;\\
\lseq \Gamma_{toleak} = \bot \land \Gamma_{toleak\_ret} = \bot,\, \Gamma_{toleak} = \bot\land \Gamma_{toleak\_ret} = \bot\rseq\\
	r0 := toleak\_ret;// \mbox{ if $\neg\df{toleak\_ret}$ then $toleak\_ret=toleak\_ret^\base$ since $toleak\_ret$ local}\\
\lseq \Gamma_{toleak} = \bot \land \Gamma_{r0} = \bot,\, \Gamma_{toleak} = \bot\land \Gamma_{r0} = \bot\rseq\\
	toleak := r0;\t1 // \mbox{This store may be bypassed}\\
\lseq \Gamma_{toleak} = \bot,\, \Gamma_{toleak} = \bot\rseq\\
	r1:= toleak;\t1 // \mbox{ if $\neg\df{toleak}$ then $toleak=toleak^\base$ since $toleak$ is local}\\
\lseq \Gamma_{r1} = \bot,\, \Gamma_{r1} = \bot\rseq\\
	\leak r1;\zbreak
\lseq true,\, true\rseq\\
	r2 := publicarray[r1*512];\\
\lseq true,\, true\rseq
\]

\subsection*{Case 12}

The code is secure since there is no store to bypass. 

\[\lseq \M ...,\, true\O\rseq\\	
	r0 := idx;\\
\lseq \M ...,\, \Gamma_{r0 \& (array\_size-1)}=\bot \land \Gamma_{publicarray[r0 \& (array\_size-1)]} = \bot\O\rseq\\
	r1 := array\_size; \\
\lseq \M ...,\, \Gamma_{r0 \& (r1-1)}=\bot \land \Gamma_{publicarray[r0 \& (r1-1)]} = \bot\O\rseq\\
	r0\_ret := r0 ~\&~ (r1-1);\\
\lseq \M ...,\, \Gamma_{r0\_ret}=\bot \land \Gamma_{publicarray[r0\_ret]} = \bot\O\rseq\\
	\fence;\\
\lseq \M ...,\, \Gamma_{r0\_ret}=\bot \land \Gamma_{publicarray[r0\_ret]} = \bot\O\rseq\\
	r0 := r0\_ret;\\
\lseq \M ...,\, \Gamma_{r0}=\bot \land \Gamma_{publicarray[r0]} = \bot\O\rseq\\
	\leak r0;\\
\lseq \M ...,\, \Gamma_{publicarray[r0]} = \bot\O\rseq\\
	r1:= publicarray[r0];\\
\lseq \Gamma_{r1} = \bot,\, \Gamma_{r1} = \bot\rseq\\
	\leak r1;\\
\lseq ...,\, true\rseq\\
	r2 := publicarray2[r1*512];\\
\lseq true,\, true\rseq
\]

\subsection*{Case 13}

The code is secure since the only store is followed by a fence (and hence cannot be bypassed).

\[\lseq \M ...,\, true\O\rseq\\	
	r0 := idx;\\
\lseq \M ...,\, \Gamma_{r0 \& (array\_size-1)}=\bot \land \Gamma_{publicarray[r0 \& (array\_size-1)]} = \bot\O\rseq\\
	r1 := array\_size; \\
\lseq ...,\, \Gamma_{r0 \& (r1-1)}=\bot\land \Gamma_{publicarray[r0 \& (r1-1)]} = \bot\rseq\\
	r0 := r0 ~\&~ (r1-1);\\
\lseq ...,\, \Gamma_{r0}=\bot\land \Gamma_{publicarray[r0]} = \bot\rseq\\
	\leak r0;\\
\lseq ...,\, \Gamma_{publicarray[r0]} = \bot\rseq\\
	r1 := publicarray[r0];\\
\lseq ...,\, \Gamma_{r1} = \bot\rseq\\
	toleak\_ret := r1;\zbreak
\lseq true,\, \Gamma_{toleak\_ret} = \bot\rseq\\
	\fence;\\
\lseq \Gamma_{toleak\_ret} = \bot,\, \Gamma_{toleak\_ret} = \bot\rseq\\
	r1 := toleak\_ret;\t1 // \mbox{ if $\neg\df{toleak}$ then $toleak\_ret=toleak\_ret^\base$ since $toleak\_ret$ local}\\
\lseq \Gamma_{r1} = \bot,\, \Gamma_{r1} = \bot\rseq\\
	\leak r1;\\
\lseq true,\, true\rseq\\
	r2 := publicarray2[r1*512];\\
\lseq true,\, true\rseq
\]

\section{Combination of Spectre-PHT and Spectre-STL}

Below we apply our logic to Listing~1 from Fabian et al.\ \cite{fab22} which involves a vulnerability arising from a combination of Spectre-PHT and Spectre-STL. The tests are reexpressed in the language from Section~\ref{sec:wp}. In the test, $secret$ is a pointer to a sensitive value, and $public$ a pointer to a non-sensitive value. $p$ is a local variable and $A$ a publicly accessible array.

The speculative precondition is elided whenever there are no points where speculation can start (stores or branches) earlier in the code.\\

\noindent To express this test in our simple programming language, pointers are modelled as arrays of length 1.

 The code is insecure since it requires that the value at $secret$ is non-sensitive. 

\[\lseq ...,\, \Gamma_{public[0]} = \bot \land \Gamma_{secret[0]} = \bot \land \Gamma_{p[0]}=\bot\rseq\\
	r0 := 0;\\
\lseq ...,\, \Gamma_{public[0]} = \bot \land \Gamma_{secret[0]} = \bot \land \Gamma_{r0}=\bot \land \Gamma_{p[0]}=\bot\rseq\\
	r1 := secret;\\
\lseq ...,\, \Gamma_{public[0]} = \bot \land \Gamma_{r1[0]} = \bot \land \Gamma_{r0}=\bot \land \Gamma_{p[0]}=\bot\rseq\\
	p := r1;\\
\lseq \M(\df{p[0]} \implies \Gamma_{p[0]} = \bot) \land (\lnot\df{p[0]} \implies \Gamma_{p[0]^{\base}} = \bot) \land \Gamma_{public[0]} = \bot,\,\\
\Gamma_{public[0]} = \bot \land \Gamma_{p[0]} = \bot \land \Gamma_{r0}=\bot\rseq\O\\
	r1 := public; \t1 // \mbox{ if $\neg\df{public}$ then $public=public^\base$ since $public$ is local}\\
\lseq \M(\df{p[0]} \implies \Gamma_{p[0]} = \bot) \land (\lnot\df{p[0]} \implies \Gamma_{p[0]^{\base}} = \bot) \land \Gamma_{r1} = \bot,\,\\
\Gamma_{r1[0]} = \bot \land \Gamma_{p[0]} = \bot \land \Gamma_{r0}=\bot\rseq\O\\
	p := r1;   \t1 //  \mbox{ This store may be bypassed}\\
\lseq \M (\df{p[0]} \implies \Gamma_{p[0]} = \bot) \land (\lnot\df{p[0]} \implies \Gamma_{p[0]^{\base}} = \bot),\, \Gamma_{p[0]} = \bot \land \Gamma_{r0}=\bot)\rseq\\
	\ifc (r0 != 0)\\ 
\t1 \M \lseq (\df{p} \implies \Gamma_{p[0]} = \bot) \land (\lnot\df{p} \implies \Gamma_{p[0]^{\base}} = \bot),\, \Gamma_{p[0]} = \bot\rseq\\
	r1 := p[0];\\
\lseq \Gamma_{r1} = \bot,\, \Gamma_{r1} = \bot\rseq\\
	\leak r1;\\
\lseq true,\, true\rseq\\
	  r2 := A[r1*512];\\
\lseq true,\, true\rseq\O\\
\lseq true,\, true\rseq
\]

\end{document}